\documentclass[aps,prx,reprint,10pt,twocolumn,citeautoscript,superscriptaddress]{revtex4-1}
\pdfoutput=1

\usepackage{natbib}
\usepackage[english]{babel}
\usepackage{letltxmacro}
\usepackage{latexsym}
\LetLtxMacro{\ORIGselectlanguage}{\selectlanguage}
\makeatletter
\DeclareRobustCommand{\selectlanguage}[1]{%
  \@ifundefined{alias@\string#1}
    {\ORIGselectlanguage{#1}}
    {\begingroup\edef\x{\endgroup
       \noexpand\ORIGselectlanguage{\@nameuse{alias@#1}}}\x}%
}
\newcommand{\definelanguagealias}[2]{%
  \@namedef{alias@#1}{#2}%
}
\makeatother
\definelanguagealias{en}{english}
\definelanguagealias{English}{english}
\usepackage{graphicx}
\usepackage{amsmath}
\usepackage{amsfonts}
\usepackage{amssymb}
\usepackage{bm}
\usepackage{color}
\usepackage[percent]{overpic}
\usepackage{soul} 
\usepackage{amssymb}
\usepackage{wasysym}
\usepackage{dsfont}
\usepackage{float}
\usepackage[mathscr]{euscript}
\usepackage{lipsum}
\usepackage{hyperref}
\hypersetup{
    pdfstartview={FitH},    
    colorlinks=true,       
    linkcolor=blue,          
    citecolor=blue,        
    filecolor=magenta,      
    urlcolor=blue           
}

\newcommand{\be}{\begin{equation}}
\newcommand{\ee}{\end{equation}}
\newcommand{\bea}{\begin{eqnarray}}
\newcommand{\eea}{\end{eqnarray}}

\newcommand{\sign}{\mathrm{sign}}

\newcommand{\mc}{\mathcal}

\newcommand{\vect}[1]{\boldsymbol{#1}}

\usepackage{graphicx}
\usepackage[colorinlistoftodos]{todonotes}
\usepackage{verbatim}
\usepackage[normalem]{ulem}
\usepackage{enumitem}

\begin{document}

\title{Quantum and classical coarsening and their interplay with the Kibble-Zurek mechanism}
\author{Rhine Samajdar}
\affiliation{Department of Physics, Princeton University, Princeton, NJ 08544, USA}
\affiliation{Princeton Center for Theoretical Science, Princeton University, Princeton, NJ 08544, USA}
\author{David A. Huse}
\affiliation{Department of Physics, Princeton University, Princeton, NJ 08544, USA}

\date{\today}

\begin{abstract}
Understanding the out-of-equilibrium dynamics of a closed quantum system driven across a quantum phase transition is an important problem with widespread implications for quantum state preparation and adiabatic algorithms. While the quantum Kibble-Zurek mechanism elucidates part of these dynamics, the subsequent and significant coarsening processes  lie beyond its scope. Here, we develop a universal description of such coarsening dynamics---and their interplay with the Kibble-Zurek mechanism---in terms of scaling theories. Our comprehensive theoretical framework applies to a diverse set of ramp protocols and encompasses various coarsening scenarios involving both quantum and thermal fluctuations. Moreover, we highlight how such coarsening dynamics can be directly studied in today's ``synthetic'' quantum many-body systems, including Rydberg atom arrays, and present a detailed proposal for their experimental observation.
\end{abstract}

\maketitle

\section{Introduction}
\label{sec:intro}

Exploring the nonequilibrium dynamics of isolated interacting quantum systems is an exciting frontier of modern condensed matter physics \cite{polkovnikov2011colloquium}. While a variety of powerful present-day analytical and numerical tools are well-equipped for investigating the dynamics of large systems for short times or small systems at long times, studying the out-of-equilibrium dynamics of large quantum systems at late times, by and large, remains a challenging problem  \cite{langen2015ultracold,mitra2018quantum}. Recently, the advent of highly controllable quantum many-body systems \cite{lloyd1996universal,buluta2009quantum,georgescu2014quantum} has opened new doors in this direction, yielding a wealth of valuable---and sometimes, surprising---insights \cite{altman2021quantum,daley2022practical}. A common operating principle in nearly every such ``analog'' system is to start in an easy-to-prepare initial state and then slowly change some parameters of the Hamiltonian to drive the system to a target quantum state; this idea, for instance, forms the basis for adiabatic quantum computing \cite{Albash.2018}. However, 
instances in which the initial and desired states are in the same quantum phase notwithstanding, such a process is fundamentally nonadiabatic because the system has to pass through a quantum phase transition (QPT) \cite{sachdev2011quantum}. Therefore, understanding the many-body dynamics across a QPT, during which the system necessarily falls out of equilibrium, is a question of both great theoretical import and practical relevance.

Generically, we are interested in a situation where a quantum system starts in the ground state of a trivial ``disordered'' phase and by changing some (dimensionless) tuning parameter of the Hamiltonian $g$, is ramped through a quantum critical point (QCP) located, without loss of generality, at $g_c$\,$=$\,$0$. We assume the existence of an ordered phase on the $g$\,$>$\,$0$ side of the QCP that persists up to some nonzero temperature or equivalently, a nonzero energy density (see Fig.~\ref{fig:outline}). As an example, one can think of a $(2$\,$+$\,$1)$D transverse-field Ising model, which undergoes a QPT from a paramagnet to a ferromagnetic phase as the ratio of the exchange coupling to the transverse field is increased. For concreteness, let us consider a protocol that starts from the ground state deep in the disordered phase, ramping with $dg/dt$\,$=$\,$1/\tau$, and stops at $g$\,$=$\,$g_s$; accordingly, $g(t)$\,$=$\,$t/\tau$, and the ramp is stopped at $t_s=\tau g_s$.  Note that we choose our zero of time so that $g(t=0)=0$; this ramp starts at $t<0$. During this process, as the system is driven into the ordered phase, the long-range correlations of the ordered phase take time to develop. 
The system forms competing and growing ``patches'' (domains) of short-range order, with defects due to the incompatibility between different such patches. The question, we then ask, is how does long-range order develop as a function of time, if it does at all?  

To address this issue, let us look at the dynamics step by step. As the ramp proceeds, initially the ground state will evolve adiabatically, building up the quantum critical correlations at short distances. On approaching the QCP, the equilibrium correlation length (in the ground state) diverges as $\xi_q$\,$\sim$\,$ l_0\, \lvert g \rvert^{-\nu}$ while the characteristic energy
scale of fluctuations above the ground state vanishes as $\Delta$\,$\sim$\,$J \,\lvert g \rvert^{\nu z}$, where $l_0$ and $J (\equiv t^{-1}_0)$ are microscopic length and energy scales, respectively \cite{sachdev2011quantum}. The correlation length exponent, $\nu$, and  the dynamical critical exponent, $z$, are determined by the universality class of the QCP. Due to this diverging relaxation time in the vicinity of the QCP, at some point, the evolution becomes nonadiabatic, as described by the quantum Kibble-Zurek mechanism (KZM) \cite{zurek2005dynamics,polkovnikov2005universal,dziarmaga2005dynamics}. Consequently, the system is driven out of equilibrium and heats up. The KZM predicts that the time at which the system becomes nonadiabatic is $t$\,$=$\,$-t_{\textsc{kz}}$, where $t_{\textsc{kz}} \sim t_0(\tau/t_0)^{\nu z/(\nu z+1)}$ and the system remains ``frozen'' through an impulse regime until time $t=+t_{\textsc{kz}}$ with a correlation length $\xi_{\textsc{kz}} \sim l_0(\tau/t_0)^{\nu/(\nu z+1)}$.  In terms of the tuning parameter, this nonadiabatic quantum critical regime is thus defined by $-g_{\textsc{kz}}<g(t)<g_{\textsc{kz}}=t_{\textsc{kz}}/\tau$. 

However, the KZM is only an approximation and in actuality, there can be significant dynamics beyond it because the system does not fully freeze when it stops following $g(t)$ adiabatically.  Instead, there is a nonequilibrium correlation length $\ell(t)$ that may continue to grow with time.
While examples of such beyond-KZM quantum dynamics have been identified \cite{roychowdhury2021dynamics,schmitt2022quantum,zeng2023universal}, their understanding so far has often been model-specific. Here, we show that these phenomena can be framed in a \textit{universal} quantum theory of coarsening. Such coarsening dynamics, which have been well-studied in classical systems \cite{furukawa1985dynamic,bray2002theory}, have recently enjoyed renewed attention in the context of phase-ordering kinetics across thermal quenches \cite{chesler2015defect} in trapped Bose gases \cite{goo2022universal,proukakis2023universality,rabga2023variations}. For quantum transitions and their associated coarsening processes \cite{chandran2013kibble,gagel2015universal}, however, much less is known, especially with regard to the interplay between quantum and thermal fluctuations.

In this work, we develop a detailed theoretical formalism to describe the various combinations of Kibble-Zurek and coarsening dynamics that can arise in the ``ramp and stop'' protocols specified above, and illustrate the experimental implications thereof. On the theoretical front, exact results have been previously obtained for quantum $O(N)$ models with $N$\,$\rightarrow$\,$\infty$ \cite{chandran2013equilibration,maraga2015aging} and for the one-dimensional transverse-field Ising spin chain \cite{kolodrubetz2012nonequilibrium}. However, these models are rather special in that they are characterized by an infinite number of conservation laws and thus do not fully thermally equilibrate, relaxing instead to a generalized Gibbs ensemble. Veering away from such nongeneric integrable models, to make theoretical progress, here we choose to work with scaling theories for domain growth and nonequilibrium critical dynamics. 
To begin, in Sec.~\ref{sec:stages}, we first outline the various types of coarsening that may be encountered by a system depending on the ramp protocol. In Sec.~\ref{sec:qcc} thereafter, we work out the domain growth laws for the simplest such case in which the system coarsens to an ordered phase. Generalizations thereof are presented in Sec.~\ref{sec:cl}, which examines the influence of thermal fluctuations near the 
critical phase boundary, and Sec.~\ref{sec:dis}, which studies a ramp that brings one to the disordered phase. Finally, in Sec.~\ref{sec:exp}, we demonstrate how these phenomena can be witnessed in today's Rydberg-atom quantum simulators and present in-depth proposals for their experimental observation, before concluding in Sec.~\ref{sec:end}.

\section{Stages of coarsening}
\label{sec:stages}

Once the system falls out of equilibrium during the course of the ramp, its correlation length $\ell(t)$ stops tracking that of the instantaneous ground state $\xi_q(g(t))$.  Since the system heats up to above the ground state, another length scale enters the problem, namely, the thermal equilibrium correlation length $\xi_{\rm{th}}$, which depends on $g$ and the energy density $\varepsilon$.
While the system is within  the quantum critical fan, it may first undergo a small amount of ``quantum critical coarsening'' such that quantum critical correlations are established to the growing length scale $\ell (t)$.  This regime applies when $\ell(t) \lesssim \rm{min}\{\xi_q,\xi_{\rm{th}}\}$.  The correlation lengths $\xi_q$ and $\xi_{\rm{th}}$ diverge as the ramp passes the quantum critical point and the phase transition, respectively.  Then, if the ramp continues, we enter the ordered phase and these two correlation lengths decrease with increasing $g$ as we go deeper into the ordered phase.  We exit the quantum critical coarsening regime when these correlation lengths, which by then are of similar magnitude, fall below the nonequilibrium correlation length $\ell(t)$ that we have grown. This puts us in the noncritical coarsening regime of the ordered phase where $\ell(t)>\xi_{\rm{th}}\geq \xi_q$ and $\ell(t)$ continues to grow with time.  This is in sharp contrast to the KZM prediction, where $\ell(t)\sim\xi_{\textsc{kz}}$ is frozen and set entirely by the freeze-in/out point, which determined $\xi_{\textsc{kz}}$.   

\begin{figure}[tb]
    \centering
    \includegraphics[width=0.9\linewidth]{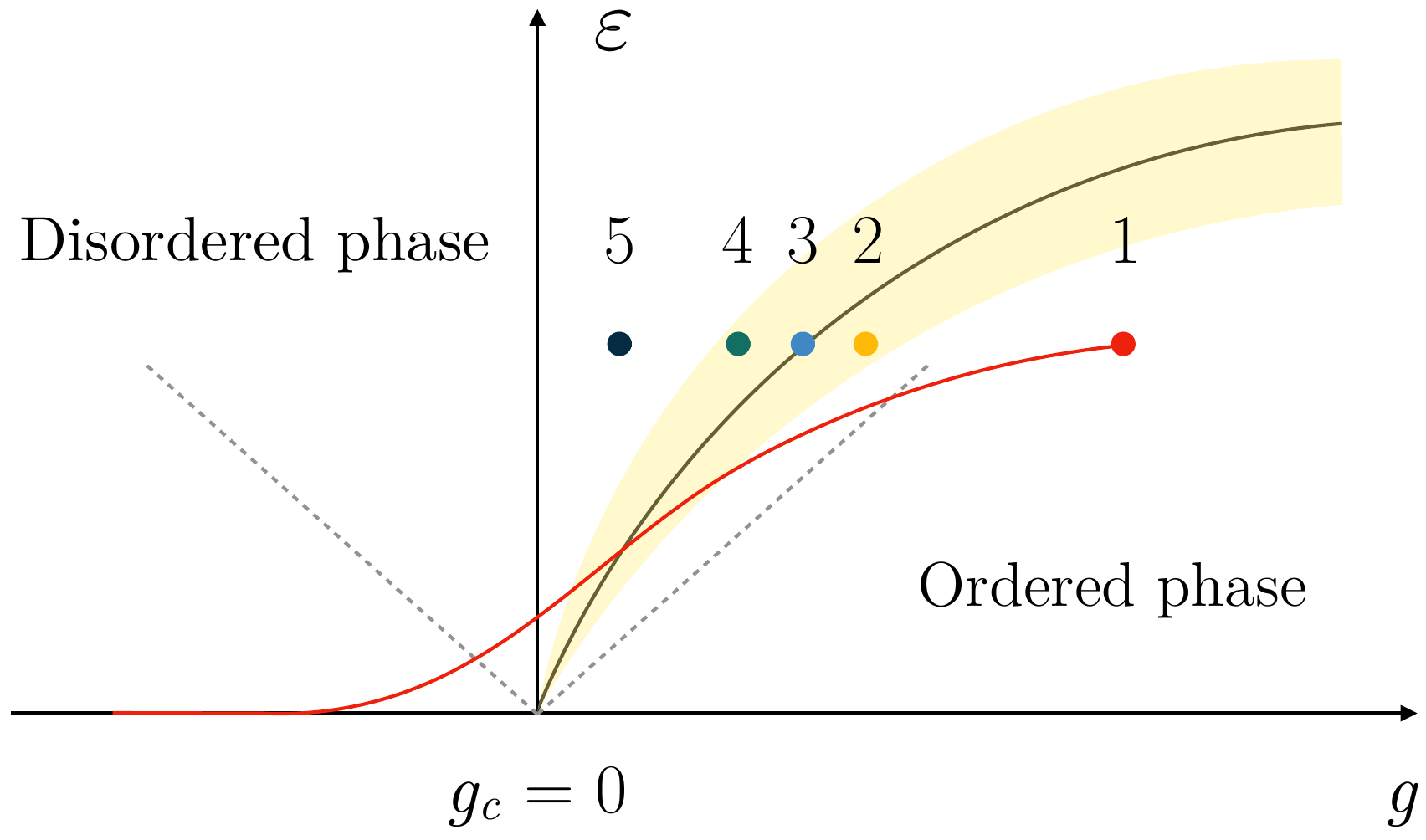}
    \caption{Schematic phase diagram as a function of the tuning parameter $g$ and the energy density $\varepsilon$. The quantum critical fan is marked by dashed grey lines while the classical critical region is shaded in yellow. A ramp across the quantum phase transition stopping well in the ordered phase is drawn in red. The numbered circles indicate representative end points for ramps leading to the different coarsening cases 1 through 5. }
    \label{fig:outline}
\end{figure}

Other scenarios can occur if we stop the ramp at $g_s$, depending on $g_s$ and on the heating which has occurred during the nonadiabatic part of the ramp. One option is to stop the ramp at a point on or near the 
phase boundary, which we assume is critical, and where $\xi_{\rm{th}}\gg\xi_q$.  In this case, after some quantum critical coarsening, the subsequent dynamics will quickly cross over 
to a regime $\xi_{\rm{th}}\gg\ell(t)>\xi_q$ of ``classical critical coarsening'' with a dynamical exponent $\bar{z}$ (here and throughout, we denote  
classical critical exponents with a bar).  If we are exactly at the critical energy density for $g_s$, then this critical coarsening can continue without limit.  On the other hand, if we are on the disordered side of the transition, the system will eventually equilibrate to the disordered phase once $\ell(t)$ reaches $\xi_{\rm{th}}$, the equilibrium correlation length.  Depending on how close we are to the transition, there may or may not be an interval of classical critical coarsening between the quantum critical regime and this final equilibration.  If we instead stop in the ordered phase, once the growing correlation length $\ell(t)$ exceeds the equilibrium correlation length $\xi_{\rm{th}}$,  we then cross over to standard noncritical coarsening with an exponent $z_d$ that depends on the dynamical model as classified, e.g., by \citet{bray2002theory}.  Again, an interval of classical critical coarsening may occur, depending on how close we are to the phase boundary.

Considering 
the scenarios outlined above, we can now identify five distinct situations according to the regimes that the system passes through over the course of its time evolution:\\
Case 1: Quantum critical + noncritical coarsening,\\
Case 2: Quantum critical + classical critical + noncritical coarsening,\\
Case 3: Quantum critical + classical critical coarsening,\\
Case 4: Quantum critical + classical critical coarsening\phantom{,} $\rightarrow$ disordered phase,\\
Case 5: Quantum critical coarsening $\rightarrow$ disordered phase.\\
For a ramp without stopping, the only possibility is case 1; the other cases occur when we stop the ramp within the quantum critical fan at $-t_{\textsc{kz}}<t_s\lesssim t_{\textsc{kz}}$.  Note that there is no scenario in which we only have quantum critical coarsening---in isolation---because the nonadiabatic ramp puts the system out of equilibrium and thus at an energy above the ground state.  Then either the system equilibrates directly to the 
disordered phase, or $\ell (t)$ grows sufficiently so that the behavior at that length scale crosses over to a classical regime of coarsening.  Note here that the regimes that we label ``classical'' could be instead labelled ``thermal'', in that they are regimes where thermal fluctuations are strong at the scale being considered.  Thus, noncritical coarsening is also a type of ``classical'' coarsening in this sense.  The microscopic dynamics are assumed to remain fully quantum in all regimes, since we are studying the unitary dynamics of an isolated quantum system.

These five cases are illustrated in the phase diagram in  Fig.~\ref{fig:outline}. Since we are considering a fundamentally out-of-equilibrium situation, it is useful to think about the phase diagram in terms of the energy density rather than the temperature. As coarsening progresses, energy is transferred from the domain walls or other defects to thermal excitations inside the domains.  While the ramp continues, the drive will in general change the energy density, but after the ramp is stopped the total energy density remains constant, since we assume the system is otherwise isolated.

\section{Quantum critical and noncritical coarsening}
\label{sec:qcc}

As a starting point, we now study in detail the first of the five cases listed above, which will then set the framework for the other four. Specifically, we are interested in a situation where the system coarsens in an ordered phase at late times but without any classical critical coarsening step. Unless specified otherwise, we work in units where the microscopic time and length scales, $t_0$ and $l_0$, are set to one.

\subsection{Indefinitely continuing ramp}
\label{sec:continue}

First, we consider a scenario in which a slow ramp continues steadily as $g(t)=t/\tau$ well into the noncritical coarsening regime of $\ell(t)>\xi_{\rm{th}}\geq \xi_q$, without stopping, with $\tau\gg t_0$. We will be interested in the growth of $\ell(t)$ both at early and late times.  
We can match the long-time scaling behavior to a model of noncritical coarsening, for which we expect that
\begin{equation}
\label{eq:expect}
 \frac{d\ell(t)}{dt} \sim \frac{\xi_q^{z_d}~\Delta}{(\ell(t))^{z_d-1}}~,
\end{equation}
as long as the ramp is slow and we remain close enough to the quantum critical point so that $\xi_q\gg l_0$; this should mean that the scaling of the rate of noncritical coarsening (i.e., $\xi_q^{z_d}\Delta$ in the numerator above) is still governed by the quantum critical scaling.  The reasoning implicit in such a growth law is that the dynamical processes varying the growth rate act on a much faster timescale than the coarsening itself \cite{biroli2010kibble}.
We are assuming power-law coarsening with $z_d<\infty$, as is generally expected unless the defects are pinned to the lattice or to quenched randomness.  The ramp is also assumed to be slow enough so that the heating is weak and the system ends up at an energy density within the ordered phase that is well below that of the phase transition.

In order to connect the different dynamical regimes, we now define a scaling ansatz for such a steady ramp, $g(t)=t/\tau$, as
\begin{equation}
\label{eq:ansatz}
    \ell(t) \approx \xi_{\textsc{kz}} f\left(\frac{t}{t_{\textsc{kz}}} \right) \equiv \xi_{\textsc{kz}} f\left(x \right),
\end{equation}
where $f(x)$ is a universal function and $\xi_{\textsc{kz}}$ and $t_{\textsc{kz}}$ depend on $\tau$, as given above. In terms of the scaling variable $x$, we can identify three regimes as
\begin{alignat}{2}
\nonumber
    &x\ll -1: &&\mbox{ adiabatic},\\
    &|x|\lesssim \mathcal{O}(1): &&\mbox{ quantum critical coarsening},\\
    \nonumber
    &x\gg1:&& \mbox{ noncritical coarsening}.
\end{alignat}
The behavior in each regime can be established as follows. When $x \ll -1$, the system is able to respond dynamically to the change in the Hamiltonian's parameters since its equilibrium relaxation time (defined by the inverse energy gap of the instantaneous ground state), $t_q$\,$\sim$\,$\vert g(t)\vert^{-\nu z}$, is smaller than $|g|/\dot{g}=|t|$. Hence, the system evolves adiabatically and stays in
equilibrium with the running $g(t)$, so in this regime $\ell(t)\approx\xi_q$, which gives 
\begin{equation}
\label{eq:f1}
    f(x) \sim |x|^{-\nu} \quad \mbox{ for }x \ll -1.
\end{equation}

Next, as the ramp passes through the quantum critical fan, there is nothing singular, since the system has heated up above the ground state, so it avoids the true quantum critical point.  The scaling function $f(x)$ remains of order one and nonsingular in this quantum critical regime:
\begin{equation}
\label{eq:f2}
    f(x) = \mc{O}(1) \quad \mbox{ for }|x| \sim \mathcal{O}(1).
\end{equation}
Correspondingly, $\ell(t)$ grows by a universal quantum critical coarsening factor of $f(+1)/f(-1)$ as the ramp passes through the quantum critical fan.  
This ratio is an $\mathcal{O}(1)$ factor that cannot be determined by scaling arguments alone. Thus, the amount of quantum critical coarsening is rather limited.

\begin{figure}[tb]
    \centering
    \includegraphics[width=0.9\linewidth]{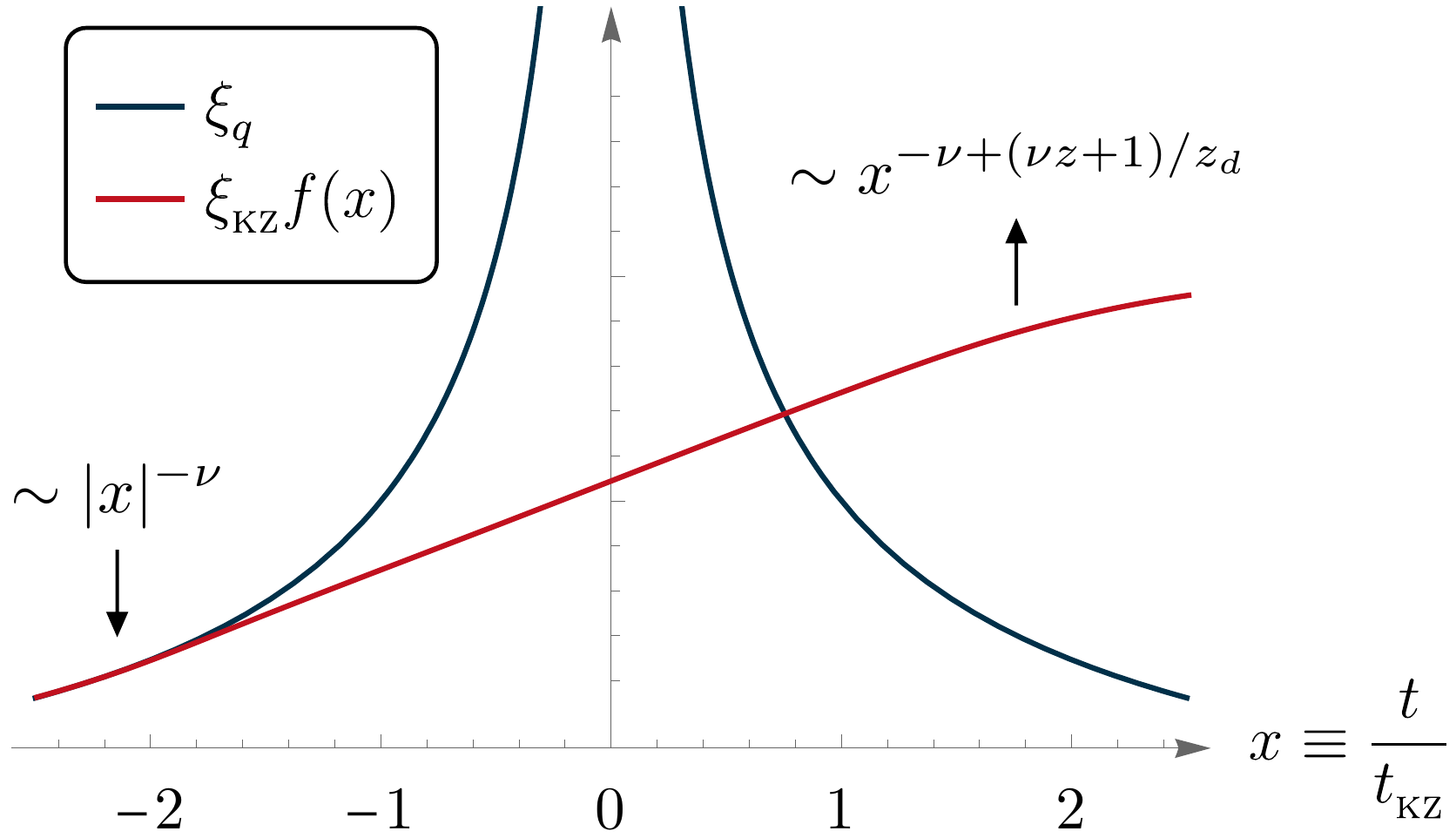}
    \caption{Schematic depiction of the universal scaling function $f(x)$  [Eq.~\eqref{eq:ansatz}] for the case where the ramp is continued indefinitely. The nonequilibrium correlation length (red) tracks $\xi_q$ (blue) in the adiabatic regime ($x\ll-1$) and grows by an $\mc{O}(1)$ factor by quantum critical coarsening ($ x \sim \mc{O}(1)$) before crossing over to the noncritical coarsening regime ($x \gg 1$).
    }
    \label{fig:fx}
\end{figure}

After the quantum critical fan, the ramp continues into the noncritical coarsening regime.  There has not been enough time to establish long-range order, so the system will have short-range order on length scales below $\ell(t)$, and defects or textures disrupting the order at length scale $\ell(t)$ and longer.  Matching Eq.~\eqref{eq:ansatz} to Eq.~\eqref{eq:expect} in this regime, we obtain, using $\xi_q\sim\xi_{\textsc{kz}}(t/t_{\textsc{kz}})^{-\nu}$ and $\Delta\sim(t/t_{\textsc{kz}})^{\nu z}/t_{\textsc{kz}}$, that the scaling function behaves as~\cite{chandran2013kibble,sicilia2007domain,biroli2010kibble}
\begin{equation}
\label{eq:f3}
    f(x) \sim x^{-\nu + (\nu z +1)/z_d}, \quad x \gg 1~.
\end{equation}
In other words, at late times, the nonequilibrium correlation length grows as
\begin{equation}
\label{eq:ell}
    \ell(t) \sim  \xi_{\textsc{kz}}\left(\frac{t}{t_{\textsc{kz}}}\right)^{-\nu + (\nu z +1)/z_d}.
\end{equation}
Note that a similar growth law for a steady ramp has also been derived for curvature-driven classical coarsening in two dimensions \cite{sicilia2007domain,biroli2010kibble}.   The universal scaling function $f(x)$ introduced in Eq.~\eqref{eq:ansatz} is thus defined by Eqs.~\eqref{eq:f1}, \eqref{eq:f2}, and \eqref{eq:f3}, and is sketched in Fig.~\ref{fig:fx}.  From Eq.~\eqref{eq:ell}, we see that for the cases where $\nu z+1 > \nu z_d$, $\ell(t)$ does grow, without limit, as a power of time in the limit of a slow steady ramp. If any systems exist with  $\nu z+1=\nu z_d$, these should have logarithmic coarsening with $f(x)\sim \log{x}$, while the coarsening will halt due to the ramp for cases where $\nu z+1<\nu z_d$.  For these latter cases, the scaling function instead behaves as: $f(\infty)-f(x) \sim x^{-\nu + (\nu z +1)/z_d}$ for $ x \gg 1$, with $f(\infty)$ finite.

To illustrate the growth laws thus obtained, as an example, let us consider a ramp across the (2+1)D transverse-field Ising model (TFIM) quantum critical point.  In this case, the (scalar) order parameter is not conserved (corresponding to so-called model C dynamics \cite{hohenberg1977theory}). The associated critical exponents are 
 $\nu\simeq 0.629$,  $z$\,$=$\,$1$, and $z_d$\,$=$\,$2$ \cite{bray1998defect}. Plugging in these values, we find that the coarsening length grows as $\ell(t) \sim t^{0.186}$, which is much slower than the $t^{1/2}$ growth governed by the dynamical exponent $z_d$ alone. Note that this slowdown is a consequence of the ramp being continued without end, and hence, $\xi_q$ continuing to decrease with time. While it is not possible to make the coarsening faster than $t^{1/2}$, as alluded to above, it is possible to slow it down even further or to stop it altogether. For instance, for a generalized ramp protocol where the system is taken through the QCP via a power-law sweep as $g(t)=\sign(t)|t/\tau|^p$, we can once again define a universal scaling ansatz 
 \begin{equation}
    \ell(t) \simeq \xi_{\textsc{kz}}  f_p\left(x \right),
\end{equation}
and exactly the same arguments as above lead to a scaling function
\begin{equation}
\label{eq:SFx}
f^{}_p(x) \sim
    \begin{cases}
    \lvert x \rvert^{-\nu},  &\mbox{for }\, x \ll -1\\
    \mc{O}(1), &\mbox{for }\, x \sim \mc{O}(1) \\
    x^{-p \nu + (p \nu z +1)/z_d}, &\mbox{for }\, x \gg 1
    \end{cases}.
\end{equation}
Then, deep in the ordered phase, the domain growth goes as $\ell\sim t^{-p \nu + (p \nu z +1)/z_d}$ as long as that growth exponent remains positive, but the coarsening will be limited if $p$ is too large and $z<z_d$. For instance, in the case of a cubic sweep ($p=3$) for the (2+1)D TFIM, the growth exponent evaluates to $-0.444$, indicating that the late-time coarsening never grows $\ell(t)$ beyond a finite limit (the limiting $\ell(t)$ does depend on $\tau$).

\subsection{Stopping the ramp}
\label{sec:stop}

Having laid the groundwork with the case of an indefinitely continuing ramp, we now consider a scenario in which the ramp is stopped at time $t = t_s$ corresponding to a value of $g=g_s =t_s/\tau$.  As an example, let us first consider stopping the ramp well into the noncritical coarsening regime, at $t_s\gg t_{\textsc{kz}}$.   
After $t_s$,
the nonequilibrium correlation length $\ell(t)$ continues to grow as specified by 
Eq.~\eqref{eq:expect}.  For $t\geq t_s\gg t_{\textsc{kz}}$, this yields 
\begin{equation}
\label{eq:offset}
    \ell(t) \approx [\ell^{z_d}(t_s)+c~(t-t_s)\xi_q^{z_d}(g_s)\Delta(g_s)]^{1/z_d}
\end{equation}
with some order-one constant $c$.

More generally, when we include stopping the ramp at $t_s$, the dynamical scaling akin to Eq.~\eqref{eq:ansatz} is
\begin{equation}
\label{eq:Fgen}
    \ell (t) \approx \xi^{}_{\textsc{kz}}\, \mc{F} \left( \frac{t}{t_{\textsc{kz}}}, \frac{t_s}{t_{\textsc{kz}}}\right) \equiv \xi^{}_{\textsc{kz}}\, \mc{F}(x,x_s)~;
\end{equation}
this scaling form will encompass all of the regimes that we consider, just by varying its two scaling arguments.  For $x\leq x_s$ this new scaling function clearly obeys $\mc{F}(x,x_s)=f(x)$.

Assuming that we stop the ramp at $x_s=t_s/t_{\textsc{kz}}\gg 1$ deep in the noncritical coarsening regime, the new $x>x_s$ regime of the universal scaling function is given by
\begin{equation}
\label{eq:SFy2}
\mc{F}(x, x_s) \approx
    x_s^{-\nu+(\nu z/z_d)}\left(\mc{C}x-\mc{C}_s x_s\right)^{1/z_d}~, 
\end{equation}
for some order-one constants $\mc{C}$\,$>$\,$\mc{C}_s$.  Note that for $z$\,$<$\,$z_d$, as is the case in many examples, the coarsening slows down as we move the stopping time deeper into the ordered phase, thereby increasing $x_s$.

We can also pick $t_s$ 
such that the ramp stops at $x_s $\,$\sim$\,$ \mc{O}(1)$, in the quantum critical regime but in the ordered phase and outside of the classical critical regime associated with the phase transition. Then, for $x$ of order one, we are still in the quantum critical coarsening regime and the scaling function will be different and of order one.  However, at long times $x\gg x_s$, the system goes into the noncritical coarsening regime where the 
scaling function behaves as
\begin{equation}
\label{eq:SFy3}
\mc{F}(x, x_s) \approx
\mc{G}(x_s)~x^{1/z_d}~; 
\end{equation}
the universal scaling function $\mc{G}(x_s)$ is of order one when $x_s$ is, and behaves as $\mc{G}(x_s)\approx \mc{C}\,x_s^{-\nu+(\nu z/z_d)}$ for $x_s\gg 1$.

\section{Intermediate classical critical coarsening}
\label{sec:cl}

\footnotetext[2]{This is because the smallest $\ell(t)$ can be is of $\mc{O} (\xi_{\textsc{kz}})$, and if we go to or past the classical transition line, then $\xi_q$ is only an $\mc{O}(1)$ factor larger than $\xi_{\textsc{kz}}$.}

The previous section examined the various scenarios in which a system proceeds directly from quantum critical to noncritical coarsening. Now, we consider the situation when there is an intermediate period of classical critical coarsening between these two regimes, which pertains to cases 2 and 3 listed in Sec.~\ref{sec:stages} above. We emphasize that in order to see the classical critical coarsening, one has to stop the ramp \textit{in} the classical critical regime.  During a ramp that instead passes through this regime without stopping, the system does not remain in this regime long enough to transition from quantum critical to classical critical coarsening. Likewise, if the ramp is stopped earlier---say, in a quantum critical coarsening regime such that $\ell(t)\ll\xi_q (g_s)$---then we cannot simultaneously also be at or past the transition into the ordered phase \cite{Note2}, and therefore, never encounter the classical critical physics. 

The key step here is to incorporate the effect of thermal fluctuations near the classical critical line within the quantum critical fan into our dynamical scaling theory. To begin, we recall that in the vicinity of a QCP, the singular part of the free energy obeys the hyperscaling form \cite{wu2011entropy}
\begin{equation*}
    F = F^{}_0\, T^{(d+z)/z} \Lambda\left(g/T^{1/(\nu z)}\right) 
\end{equation*}
for some universal function $\Lambda$, where $T$ is the temperature. Since $T$ scales as $\xi_q^{-z}$, this implies that $F $ scales as $ \xi_q^{-(d+z)}$. Recognizing that $F$\,$=$\,$E$\,$-$\,$TS$, we can use the fact that the  dominant contribution to $E$ is from the defect density on the scale $\xi_{\textsc{kz}}$ frozen in at $t_{\textsc{kz}}$ to estimate the excess energy density above the ground state as \cite{de2010quench}
\begin{equation}
\label{eq:q}
    \varepsilon \sim \xi_{\textsc{kz}}^{-(d+z)}.
\end{equation}

First, let us consider the thermal equilibrium correlation length along the path taken by the ramp, $\xi_{\mathrm{th}}$, for which we expect a scaling relation of the form
\begin{alignat}{1}
\label{eq:xicl}
\nonumber\xi^{}_{\mathrm{th}} &\simeq \xi^{}_q (g(t))\,\tilde{h} \left(\xi^{}_q(g(t))/\xi^{}_{\textsc{kz}}\right) \\
&= \xi^{}_{\textsc{kz}} \,h \left(g(t)/ g_{\textsc{kz}}\right) \equiv \xi^{}_{\textsc{kz}} \,h (y),
\end{alignat}
where $g_{\textsc{kz}}$\,$\equiv$\,$\xi_{\textsc{kz}}^{-1/\nu}$; note that $y = \mathrm{min}(x, x_s)$.
Then, as $\xi^{}_{\textsc{kz}}$ is related to the energy density by Eq.~\eqref{eq:q}, $h (y)$ encodes the classical critical divergence at some particular value of the scaling variable $y_c$.  The behavior of the function $h(y)$ is as follows:
\begin{equation}
\label{eq:fcl}
h(y) \sim
    \begin{cases}
     |y|^{-\nu}, &\mbox{for }y \ll -1 \\   
     \mathrm{smooth}, &\mbox{for }y ~\mbox{near }0 \\   
     \lvert y - y_c \rvert^{-\bar{\nu}}, &\mbox{for }y ~\mbox{near}~y_c \\  
     y^{-\nu}, &\mbox{for }y \gg 1 
    \end{cases};
\end{equation}
we now explain each of these four pieces individually.
The first case ($y$\,$\ll$\,$-1$) describes the adiabatic portion of the ramp, during which $\xi_{\mathrm{th}}$ simply mirrors $\xi_q$. The second ($y$ near 0) corresponds to the quantum critical coarsening regime, where we know previously, from Eq.~\eqref{eq:f2}, that there should not be any singularity since the ramp positions us at a nonzero energy density. Next, in the classical critical regime ($y~\mbox{near}~y_c$), the singularity associated with the divergence of the thermal equilibrium correlation length is also manifest in $h(y)$. The last case, which is perhaps the most subtle, describes the regime of long times compared to the quantum equilibration time. As we go farther into the ordered phase, the energy per length of the domain walls increases as $\xi_q^{-d}$ whereas the energy per excitation within the domains grows like the gap as $\xi^{-z}_q$. The coarsening steadily moves energy from the walls to the within-domain excitations, but at the same time, the domain wall energies also grow faster ($d$\,$>$\,$z$).  If the excitations within the domains remain dilute (i.e., with density $\ll \xi_q^{-d}$), then we should have $\xi_{\mathrm{th}}\simeq\xi_q$ once we ramp deep into the ordered phase; this informs our choice for the scaling of $h(y)$ when $y$\,$\gg$\,$1$.

Motivated by the structure of Eq.~\eqref{eq:ansatz}, we could, in principle, write down the coarsening length as a product of two separate functions
\begin{equation}
    \ell (t) \simeq \xi^{}_{\textsc{kz}}\, h \left(\frac{g(t)}{g_{\textsc{kz}}} \right) \tilde{\mc{F}} \left( \frac{t}{t_{\textsc{kz}}}, \frac{t_s}{t_{\textsc{kz}}}\right),
\end{equation}
where the first one encodes the equilibrium correlation length in Eq.~\eqref{eq:xicl} and the second accounts for the dynamical scaling. 
However, due to the divergence of $h$ at $y=y_c$, this would require the introduction of some (artificial) singular behavior in $\tilde{\mc{F}}(x,x_s)$ at the classical critical point to keep $\ell(t)$ finite. Therefore, to avoid this complication, the best way to write the scaling of $\ell(t)$ is via the combined universal scaling function introduced earlier, in Eq.~\eqref{eq:Fgen}, as
$\ell(t)$\,$\sim$\,$ \xi^{}_{\textsc{kz}}\, \mc{F} \left(x,x_s\right)$.

Let us next consider cases 2 and 3, where the ramp is stopped at or near the transition, with $x_s\geq x_c$.  Then, the different regimes of the function $\mc{F}$ relevant to these particular ramp protocols are
\begin{equation}
\label{eq:CCCy}
\mc{F}(x, x_s) \sim
    \begin{cases}
    \lvert x \rvert^{-\nu},  &\mbox{for }\, x \ll -1 <x_s\\
    \mc{O}(1), &\mbox{for }\, x_s\gtrsim x \sim \mc{O}(1) \\
    x^{1/\bar{z}},&\mbox{for }\, x_c\cong x_s\ll x\ll x^* (x_s)\\
    \displaystyle
    (x^*)^{\frac{1}{\bar{z}}}\left(\frac{x}{x^*}\right)^{\frac{1}{z_d}}, &\mbox{for }\, x \gg x^*(x_s),~x_s > x_c 
    \end{cases},
\end{equation}
with $x^*(x_s)\sim |x_s-x_c|^{-\bar{\nu}\bar{z}}$.
We can thus encapsulate quantum critical (second line), classical critical (third line), and standard noncritical (fourth line) coarsening in a \textit{single} scaling function. 
The value of $x^*$, at the crossover between the last two regimes, is set by the stopping point of the ramp, and it diverges if the ramp is halted right on the classical transition line, $x_s = x_c$.
Note that our analysis here automatically addresses case 3 in our list of scenarios, i.e., when the system only undergoes classical critical coarsening but no noncritical coarsening. Since this occurs only if the ramp is stopped at just the right critical energy density, $\ell(t)$ is then given by Eq.~\eqref{eq:Fgen} with $\mc{F}(x,x_s)$ the same as in Eq.~\eqref{eq:CCCy} but with the last line deleted, or equivalently, with $x^* \to \infty$.

\section{Final state in the disordered phase}
\label{sec:dis}

Lastly, we can also consider a scenario in which the ramp is stopped before it reaches the phase boundary, so it is stopped within the quantum critical fan but in the disordered phase.
Then, the system initially quantum critical coarsens but eventually passes into the disordered phase, whereafter it does not coarsen further. Such a process can additionally involve an intermediate stage of classical critical coarsening;  if so, then, per the arguments outlined in Sec.~\ref{sec:cl}, we have to necessarily stop the ramp in the classical critical coarsening regime. This particular situation corresponds to case 4 above, and the nonequilibrium correlation length is described by Eq.~\eqref{eq:Fgen} with the universal dynamical scaling form
\begin{equation}
\label{eq:disorder}
\mc{F}(x, x_s) \sim
    \begin{cases}
    \lvert x \rvert^{-\nu},  &\mbox{for }\, x \ll -1 <x_s\\
    \mc{O}(1), &\mbox{for }\, x_s \gtrsim x \sim \mc{O}(1) \\
    x^{1/\bar{z}}&\mbox{for }\, x_s<x_c\ll x\ll x^* (x_s)\\
    (x_c-x_s)^{-\bar{\nu}}, &\mbox{for }\, x \gg x^*(x_s),~x_s < x_c 
    \end{cases}.
\end{equation}
In the disordered phase, there is obviously no long-range order to coarsen to, so once the system ceases to be critical, the coarsening stops, as is reflected in the last regime of $\mc{F}(x,x_s)$ above.

Similarly, for case 5 where the ramp is stopped in the quantum critical fan but not in the classical critical regime, there is no further coarsening after the initial quantum critical part. In this regime, the scaling function behaves as 
\begin{equation}
\mc{F}(x, x_s) \sim
    \begin{cases}
    \lvert x \rvert^{-\nu},  &\mbox{for }\, x \ll -1 <x_s\\
    \mc{O}(1), &\mbox{for }\, x \gtrsim \mc{O}(1) \\
     \end{cases}.
\end{equation}
Finally, one can stop the ramp so early that the system is still adiabatic and in its ground state, in which case the applicable regimes of $\mc{F}(x, x_s)$ are
\begin{equation}
\mc{F}(x, x_s) \sim
    \begin{cases}
    \lvert x \rvert^{-\nu},  &\mbox{for }\, x\leq x_s \ll -1\\
    |x_s|^{-\nu}, &\mbox{for }\, x \geq  x_s
    \end{cases},
\end{equation}
with $x_s \ll x_c$.  

We emphasize that throughout our description of the different coarsening scenarios, there is just the one scaling function, $\mc{F}(x, x_s)$, and in all our discussions here, we have listed and described the many regimes of its behavior.

\begin{figure*}[tb]
\includegraphics[width=\linewidth]{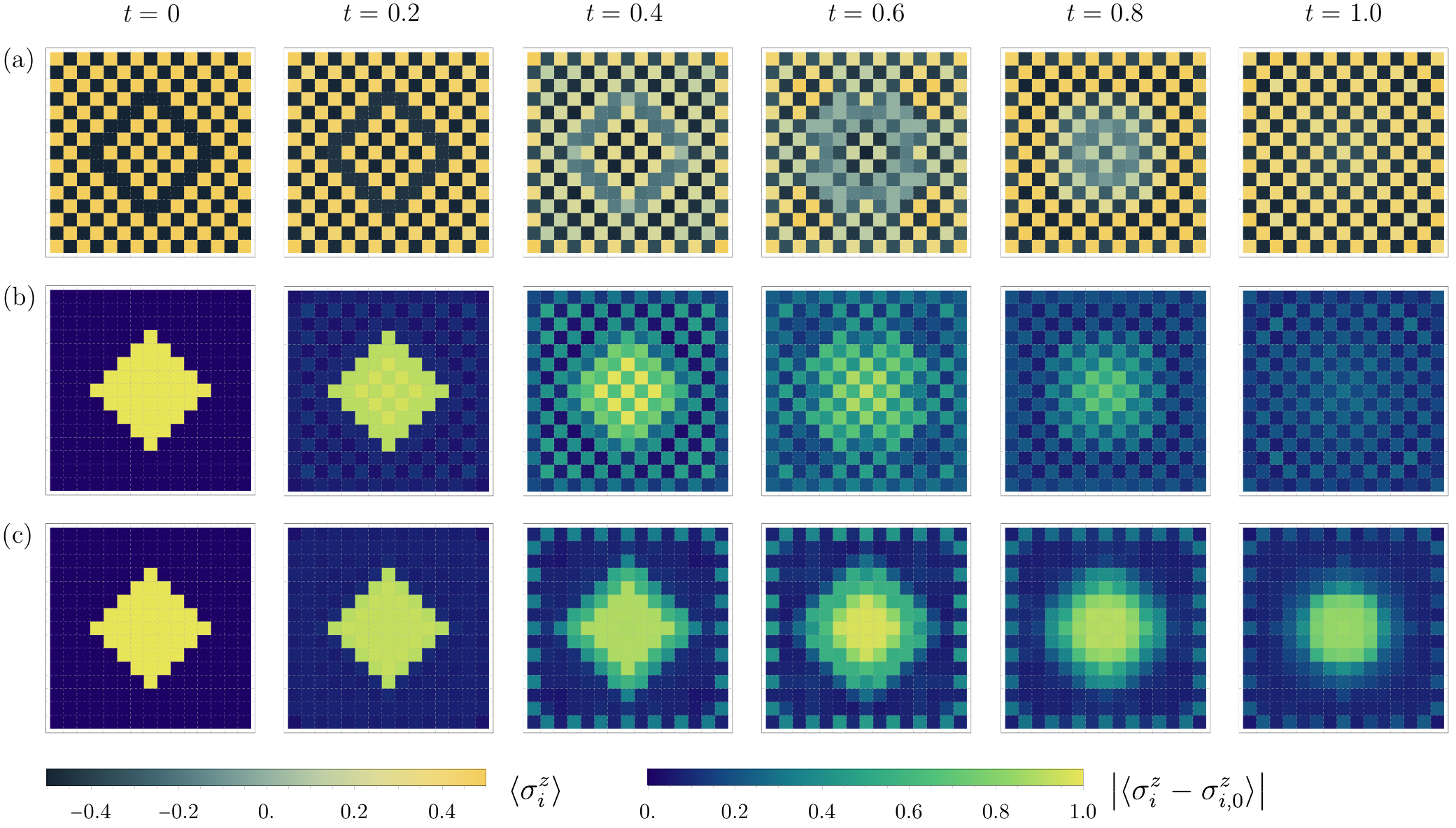}
\caption{(a) Coarsening dynamics of the spin texture $\langle \sigma^z_i\rangle$, starting from an initial domain-wall state at $t$\,$=$\,$0$ that evolves with the Hamiltonian~\eqref{eq:HRyd} for $\Delta/\Omega$\,$=$\,$2.0$ and $R_b/a$\,$=$\,$1.1$. (b) Same as in (a), but with a background configuration $\langle \sigma^z_{i,0}\rangle$---corresponding to a perfect antiferromagnetic state for the $15$\,$\times$\,$15$ lattice---subtracted out, thus demarcating domains with contrasting symmetry-breaking orders. (c) Same as in (b) but for time evolution under $H$ with $\Delta/\Omega = 4.0$.}
\label{fig:num}
\end{figure*}

\section{Experimental implementation}
\label{sec:exp}

The dynamics arising from the combination of the quantum KZM and coarsening, as formulated in the previous sections, can be observed on a wide variety of quantum simulators, which routinely employ ramps across QPTs to prepare desired target states. Here, we focus on one such platform, namely, Rydberg atom arrays \cite{endres2016atom,Browaeys.2020,morgado2021quantum}, and present a concrete proposal to demonstrate and quantify such coarsening phenomena.

Specifically, we consider a two-dimensional array of neutral atoms trapped in optical tweezers (with interatomic spacing $a$), in which each atom can either be in the ground state $\rvert g\rangle$ or a highly excited Rydberg state $\rvert r \rangle$ with a large principal quantum number. An individual atom can thus be regarded as a two-level (spin) system, and the interacting ensemble thereof can be mapped (with $|g\rangle=\vert\downarrow\rangle$ and $|r\rangle=\vert\uparrow\rangle$) to a long-ranged 2D transverse-field Ising model in a longitudinal field, described by the many-body Hamiltonian \cite{Ebadi.2021, scholl2021quantum}
\begin{equation}
\label{eq:HRyd}
H = \frac{\Omega}{2} \sum_i \sigma^x_i - \Delta \sum_i n^{}_i  + \sum_{i<j} V^{}_{ij} n^{}_i n^{}_j,
\end{equation}
where $n_i \equiv (\sigma^z_i + 1)/2$.
Here, $\Omega$ denotes the Rabi frequency characterizing the coherent oscillations between $\rvert g\rangle$ and $\rvert r\rangle$, $\Delta$ is the laser detuning, and $V_{ij} = V_0/\lvert\vect{r}_i-\vect{r}_j\rvert^6$ is the van der Waals interaction between two excited atoms that are spatially located at positions $\vect{r}_i, \vect{r}_j$. The strong Rydberg-Rydberg interactions lead to the ``Rydberg blockade'' effect \cite{jaksch2000fast} in which the excitation of one atom to the Rydberg state prevents the simultaneous excitation of another nearby atom. The strength of the interactions---which can be controlled by varying the distance between adjacent tweezers---can thus be conveniently parametrized in terms of the Rydberg blockade radius, defined as $R_b \equiv (V_0/\Omega)^{1/6}$. 

In recent years, such neutral atom arrays have emerged as versatile platforms to study a variety of strongly correlated phases of matter \cite{de2019observation,Semeghini.2021,Samajdar.2021,Verresen.2020,PhysRevLett.130.043601,chen2023continuous}, quantum phase transitions \cite{keesling2019quantum,samajdar2018numerical,whitsitt2018quantum,PhysRevLett.122.017205,chepiga2021kibble}, and nonequilibrium quantum dynamics \cite{bernien2017probing,turner2018weak,bluvstein2021controlling,hannes_dynamical}. In particular, if the atoms are arranged in a square array such that nearest neighbors are blockaded, the simplest nontrivial many-body ground state (for sufficiently large $\Delta/\Omega$\,$>$\,$0$) is an antiferromagnetically ordered checkerboard (or N\'eel) state \cite{Samajdar_2020,kalinowski2021bulk}. Experimentally, such a checkerboard state can be prepared by initializing all atoms in the atomic ground state $\rvert g \rangle$ at $\Delta/\Omega < 0$ (which corresponds to a trivial disordered phase), and then ramping up $\Delta/\Omega$ to large positive values \cite{Ebadi.2021,Kim.2021}. The quantum phase transition encountered in this process between the disordered and checkerboard phases belongs to the (2+1)D quantum Ising universality class.

To observe the coarsening dynamics, we propose two kinds of experiments. In the first type, the Hamiltonian is tuned across the aforementioned phase transition by varying $\Delta/\Omega$ from negative to positive values at various speeds $1/\tau$. The system is (destructively) imaged at different points along the ramp schedule---corresponding to different final detunings---to measure the nonequilibrium correlation length. This gives us access to the growth of $\ell(t)$ for the case of an indefinitely continuing ramp described in Sec.~\ref{sec:continue} and is similar in spirit to the experiments carried out by \citet{Ebadi.2021}. Note that the ramp speed also determines the relative extents of quantum critical and noncritical coarsening during the dynamics. Alternatively, one can sweep to different final detunings, stop the ramp, and hold for variable times $t_{\mathrm{hold}}$ before measuring the correlation length. When the ramp is stopped well into the ordered phase and $t_{\mathrm{hold}}$ is large relative to the duration of the ramp itself, one should be able to observe the increase of $\ell (t) \sim t^{1/2}$, as predicted in Sec.~\ref{sec:stop} for case 1 at long times.  
One can also access cases 2 through 5 by varying where the ramp is stopped, and also possibly varying $1/\tau$.

\begin{figure}[tb]
\includegraphics[width=\linewidth]{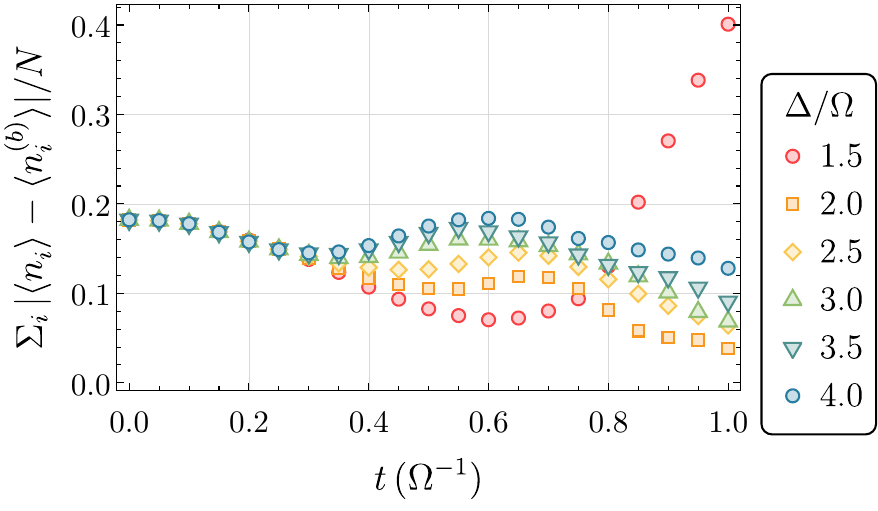}
\caption{Site-averaged difference between the Rydberg excitation densities during the time evolution starting from a domain-wall initial state, $\langle n_i \rangle$, and an ideal background  antiferromagnetic configuration, $\langle n^{(b)}_i \rangle$, for various values of $\Delta/\Omega$, illustrating that the coarsening slows down as one progresses deeper into the ordered phase.}
\label{fig:numb}
\end{figure}

The second class of experiments that we envision begin by preparing a deterministic domain-wall initial state, which has been recently enabled by the capability of locally controlling individual qubits. To be specific, we use local light shifts to imprint a specific mask of initial Rydberg excitations on the array. Then, the interacting Hamiltonian $H$ is turned on, which corresponds to a quench to a specific final detuning, and one follows the relaxation of the domain wall(s) as the coarsening dynamics progress. To demonstrate this process, in Fig.~\ref{fig:num}, we numerically simulate the time-evolution dynamics starting from an initial state with a single large domain wall built by embedding a large antiferromagnetically ordered domain within another with the opposite antiferromagnetic order. 
For our simulations, we employ a tensor-network algorithm based on the time-dependent variational principle (TDVP) \cite{haegeman2011time,haegeman2016unifying} and maintain a truncation error of $<10^{-7}$ throughout by adaptively adjusting the bond dimension of the matrix product state up to a maximum of 4800. Our numerics are performed on a $15\times15$ array, using a blockade radius of $R_b/a$\,$=$\,$1.1$, retaining up to second-nearest-neighboring interactions (which suffices for the physics of the checkerboard phase).   In equilibrium, the quantum critical point between the disordered and the checkerboard phase is located at $\Delta/\Omega \simeq 1.0$ for these parameters  \cite{Samajdar_2020,kalinowski2021bulk}. From Fig.~\ref{fig:num}, we observe that, in consistency with our theoretical predictions, the coarsening is much faster closer to the critical point (Fig.~\ref{fig:num}(a,b); $\Delta/\Omega$\,$=$\,$2.0$), and slows down appreciably deeper in the ordered phase (Fig.~\ref{fig:num}(c); $\Delta/\Omega$\,$=$\,$4.0$), as predicted by Eq.~\eqref{eq:SFy2}. This effect, which highlights the distinction between quantum critical and noncritical coarsening, should also be observable experimentally. 
Furthermore, in Fig.~\ref{fig:numb}, we plot, as a function of time, the difference between the on-site Rydberg excitation densities during the evolution starting from a domain-wall initial state and a perfect ``background''  antiferromagnetic configuration. Once again, we see that the excess excitation density relaxes faster for lower values of $\Delta/\Omega$. Interestingly, for $\Delta/\Omega$\,$=$\,$1.5$, we find that this difference instead grows with time, indicating that the initial energy density in the domain-wall configuration is too large to allow coarsening to an ordered state. Systematically varying the size of the initially prepared domain wall---to check whether the system coarsens to an ordered phase or not---could thus be used as a thermometry protocol to experimentally map out the critical phase boundary at finite energy densities.

\section{Conclusions}
\label{sec:end}

In summary, we present a \textit{universal} description of the nonequilibrium dynamics of a generic system that is ramped across a quantum phase transition by continuously varying some tuning parameter. While the celebrated Kibble-Zurek mechanism has conventionally informed our understanding of these dynamics, our work examines and highlights the nontrivial coarsening phenomena beyond its reach that are responsible for the formation of long-range order. Specifically, we classify all the possible dynamical scenarios which may occur, depending on the ramp protocol, and systematically develop detailed scaling theories that address the different regimes of quantum critical, classical critical, and noncritical coarsening. This comprehensive framework encapsulates both the Kibble-Zurek and the coarsening dynamics in a unified fashion, offering a more complete view of  the intricate nonequilibrium dynamics that unfold during and after a quantum phase transition. 

In the broader theoretical landscape, these coarsening dynamics can also be related to the concept of nonthermal fixed points \cite{schmied2019non}. Given the power-law growth with time (of the nonequilibrium correlation length) that characterizes coarsening, the dynamics are naturally scale-invariant. This is akin to the familiar universal behavior in the vicinity of a quantum critical point where thermodynamic quantities exhibit power-law scalings with the tuning parameter, with the corresponding exponents dictated by the relevant renormalization-group fixed point. Phenomenologically, this analogy prompts the interpretation of the coarsening dynamics as a manifestation of a nonequilibrium dynamic criticality, governed by a nonthermal fixed point on the way to equilibrium.

From a practical viewpoint, our work provides useful insights for reliably preparing long-range-ordered quantum states, which lies at the heart of various quantum adiabatic algorithms and quantum simulation experiments. In particular, we have  shown how distinct types of coarsening phenomena can be directly probed in one such analog quantum simulator, namely, Rydberg atom arrays. Our analysis illustrating the importance of coarsening processes---and the dynamical signatures thereof---in this system suggests that it could be interesting to reexamine earlier experimental data \cite{keesling2019quantum, Ebadi.2021} for which the nonequilibrium dynamics across a quantum critical point were ascribed entirely to Kibble-Zurek phenomena. However, our results also provide a
solid foundation for predicting and interpreting post-transition dynamics in other quantum systems, ranging from trapped ions \cite{cui2016experimental} to Bose gases \cite{prufer2018observation}, and serve as a guide to future experiments.

\begin{acknowledgments}

We thank Sepehr Ebadi, Sarang Gopalakrishnan, Sophie Li, Mikhail Lukin, and Tom Manovitz for useful discussions. R.S. is supported by the Princeton Quantum Initiative Fellowship.  D.A.H. is supported in part by NSF QLCI grant OMA-2120757.  The numerical calculations presented in this paper were performed using the ITensor library \cite{ITensor} on computational resources managed and supported by Princeton Research Computing, a consortium of groups including the Princeton Institute for Computational Science and Engineering (PICSciE) and the Office of Information Technology's High Performance Computing Center and Visualization Laboratory at Princeton University.

\end{acknowledgments}

\bibliographystyle{apsrev4-2}
\bibliography{refs.bib}

\end{document}